# AFDI: A Virtualization-based Accelerated Fault Diagnosis Innovation for High Availability Computing

**Ameen Alkasem[1], Hongwei Liu1, Zuo Decheng[1], Yao Zhao[1]**

[1]School of Computer Science and Technology, Harbin Institute of Technology, Harbin, 150001, China

**Abstract**

Fault diagnosis has attracted extensive attention for its importance in the exceedingly fault management framework for cloud virtualization, despite the fact that fault diagnosis becomes more difficult due to the increasing scalability and complexity in a heterogeneous environment for a virtualization technique. Most existing fault diagnoses methods are based on active probing techniques which can be used to detect the faults rapidly and precisely. However, most of those methods suffer from the limitation of traffic overhead and diagnosis of faults, which leads to a reduction in system performance. In this paper, we propose a new hybrid model named accelerated fault diagnosis invention (AFDI) to monitor various system metrics for VMs and physical server hosting, such as CPU, memory, and network usages based on the severity of fault levels and anomalies. The proposed method takes the advantages of the multi-valued decision diagram (MDD), A Naïve Bayes Classifier (NBC) and virtual sensors cloud to achieve high availability for cloud services.

## 1. Introduction

With the ability to scale computing resources on demand and provide a simple pay-as-you-go business model for customers, cloud computing is emerging as an economical computing paradigm and has gained much popularity in the industry. Currently, a number of big companies such as Netflix and Foursquare have successfully moved their business services from the dedicated computing infrastructure to Amazon Elastic Computing Cloud (EC2), which is a leading public Infrastructure-as-a-Service (IaaS) cloud platform worldwide [2].

Undoubtedly, more individuals (tenants) and enterprises will leverage the cloud to maintain or scale up their business while cutting down on their budgets as reported by the International Data Corporation (IDC) which suggests that the business revenue brought by cloud computing will reach $1.1 trillion by 2015 [1].

Unfortunately, the running performance of virtual machines (VMs) on the IaaS cloud platform is unpredictable. This is due to complications in fault detection and diagnosis because of the increasing complexity and scalability in a heterogeneous environment for virtualization and clouds which present an automated and dynamic model, whereby there are conflicts with the manual/semi-automatic process of fault detection. For example, clouds are a multi-system state (MSS) consisting of multi-tenant, and multiple virtual machines (VMs) which are deployed on the same physical server. Since resources like (CPU, memory, network, etc) are not virtualized, traditional methods cannot be used to detect the kinds of faults/anomalies for these resources, such as CPU utilization, memory leak and slow hard-disk I/O. This is due to information from the VMM level and physical server resources. The reason for a decline in the performance of services or complete stop of services is due to a serious fault from one of the resources which will lead to an entire system breakdown, as shown in Fig.1.

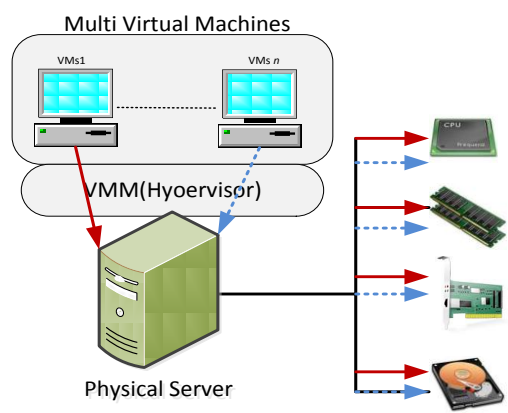

Fig.1 VMs collocated on a physical server

To address this problem, we proposed a new hybrid model to accelerated fault diagnosis framework called AFDI.

The main contributions of this paper are as follows: an AFDI model is a new approach to accelerated fault diagnosis by combining a NBC and MDD intelligent models with virtual sensors cloud to provide a decision-practical strategy for fault failure recovery of IaaS based on severity levels and anomalies, to fulfill the fault diagnosis requirements of high efficient, accuracy and learning ability.

The rest of this paper is organized as follows. Related works are in Section 2. The background of Naïve Bayes classifier, multi-valued decision diagram and virtual sensors cloud are given in Section 3. The system architecture for AFDI is presented in Section 4. Case studies of analytic models are given in section 5. The results and discussions are presented in Section 6. Finally Section 7 concludes the paper and gives some future works.

## 2. Related Works

Fault diagnosis and dynamic remediation is a popular research area. In recent years for cloud computing and intelligent models of machine learning technologies are widely used in many research and different applications in dependability. Especially for fault diagnosis and monitoring systems. In [16] authors proposed a method for the analysis of the multi-state system (MSS) structure function by a multiple-valued decision diagram. It is an effective approach for analysis and estimation of the function of high dimension in multiple-valued logic. In [19] authors presented an innovative diagnostic method by incorporating a Bayesian belief network (BBN) technique. The proposed method is capable of guiding vehicle diagnostics in a probabilistic manner. In [30] authors proposed a novel fault troubleshooting decision method based on Bayesian network and the multicriteria decision analysis. Sharma, Bikash, et al [20] they innovated CloudPD model which is the first end-to-end fault management system that can detect, diagnose, classify and suggest remediation actions for virtualized cloud-based anomalies. In [17, 18] authors proposed a high availability model in a virtualized system. This model was constructed on two non-virtualized and virtualized hosts' system models. In cloud computing, virtualized systems are distributed over different locations (heterogeneous environment). In [32] authors proposed a predictive model to predict monitored parameters from cloud infrastructure logs. In [24] authors proposed only a fault detection framework technique for virtualized environments without any suggestions for diagnosis or failure recovery.

A comprehensive analysis indicates that these researchers in different areas use Bayes' theorem only for reasoning and calculating failure rate of components. None of works took priori for the severity of the fault or full dynamical remediation for IaaS platform. This paper proposes a new approach to accelerated fault diagnosis by combining a NBC and MDD intelligent models with virtual sensors to provide a decision-practical strategy for troubleshooting of IaaS, to fulfill the fault-diagnosis requirements of high efficiency, accuracy and learning ability.

# 3. Background

Virtualization technology has been widely used today because of the prevalence of cloud computing. This section provides an overview of NBC, MDD and virtual sensors which have been used for accelerated diagnosis in virtualization environments.

## 3.1 A Naive Bayes Classifier Algorithm

A Naive Bayes classifier is a simple probabilistic classifier based on applying Bayes' theorem with independence assumptions between predictors. The Naive Bayes model is easy to build, with no complicated iterative parameter estimations which makes it particularly useful for every large datasets. In Fig.2 shows the structure of Naive Bayes graphically. It contains two kinds of nodes: a class node and attribute nodes. In Naive Bayes, though each attribute node has one class node as its parent, there are no direct links among attributes [4, 5].

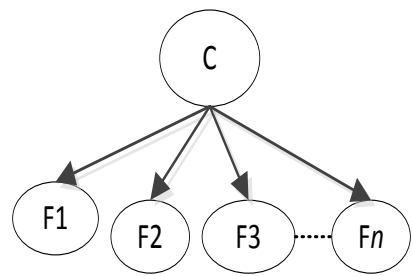

Fig.2 An example of the Naive Bayes

Despite its simplicity, the Naive Bayes classifier often outperforms more sophisticated classification methods.

### 3.1.1 The Naive Bayes Probabilistic Model

A Bayes classifier [3] is a function that maps input feature vectors $X = \{x_1, ..., x_n\}, n \geq 1$. to output class labels $C = \{c_1, ..., c_n\}$, where $X$ is the feature space. Abstractly, the probability model for Bayes classifier is a conditional model $(C|x_1, ..., x_n)$, which is the so-called posterior probability distribution. Applying Bayes' rule, the posterior can be expressed as:

$$P(C|X) = \frac{P(C, X)}{P(X)} = \frac{P(X|C)P(C)}{\sum_{i=c_i}^{c_n} P(X|c_i)P(c_i)}$$

$$= P(x_1|C) \times P(x_2|C) \times ... \times P(x_n|C) \times P(C) \quad (1)$$

Where
$P(C|X)$ : is the posterior probability of the class (target) given predictor (attribute).
$P(C)$ : is the prior probability of the class.
$P(X|C)$ : is the likelihood probability of the given class.
$P(X)$ : is the prior probability of the predictor (evidence).
Using Bayesian probability terminology, the above equation can be written simply as:

$$Posterior = \frac{prior \times likelihood}{evidence}$$

The simplest classification rule is used to assign an observed feature vector $X$ to the class with the maximum posterior probability.

$$classify(X) = arg_{c_i} max P(c_i|X)$$

$$= arg_{c_i} max \frac{P(X|c_i)P(c_i)}{\sum_{i=c_i}^{c_n} P(X|c_i)P(c_i)} \quad (2)$$

Because $\sum_{i=c_i}^{c_n} P(X|c_i)P(c_i)$ is independent of $c_i$ and does not influent the arg max operator, classify(X) can be written as

$$classify(X) = arg_{c_i} max (X|c_i) P(c_i) \quad (3)$$

This is known as the maximum a posterior (MAP) decision rule.
The major advantage of using A Bayesian networks (BN) is its ability to represent and hence understand knowledge, because BN is able to produce probability estimates rather than predictions. These estimates allow predictions to be ranked and their expected costs to be minimized. The Naive Bayes classifier still gives us a high degree of accuracy.

### 3.1.2 Confusion Matrix

One of the most common and basic ways of evaluating Naive Bayes model's performance is by creating a confusion matrix (contingency table analysis (Table 1)) and computing various metrics such as accuracy, precision, recall and so on. We can summarize large datasets and understand the measures by using NBC.

Table1: A contingency table analysis

| Prediction | Class | Observation | |
|---|---|---|---|
| | | Correct_state | Not_Correct_State |
| | Select | Ture_Positive (TP) Hit success | False_Positive(FP) False alarm type‖Error |
| | Not select | False_Negative(FN) Miss type‖Error | True_Negative(TN) Correct negative |

We utilize the following four statistical measures to evaluate the effectiveness of method by AFDI models base on Naive Bayes classifier algorithm's contingency table:

$$Rcall = \frac{TP}{TP + FN} \qquad (4)$$

$$precision = \frac{TP}{TP + FP} \qquad (5)$$

$$Accuracy = \frac{TP + TN}{TP + TN + FP + FN} \qquad (6)$$

$$False\ alarm\ rate = \frac{Number\ of\ false\ alarms}{Total\ number\ of\ alarms}$$
$$= 1 - presision \qquad (7)$$

## 3.2 Multiple-Valued Decision Diagram (MDD)

Multiple-valued decision diagrams (MDDs) are one of the effective mathematical methods for the representation of multiple-valued logic (MVL) functions of large dimensions [16]. The multiple-valued decision diagram (MDD) is a natural extension of the binary decision diagram (BDD) to a multiple-valued case [11]. The MDD is a directed acyclic graph (DAG) with up to $n$ sink nodes. The MDD model has three sink nodes, representing the corresponding component being in each of the multiple states (operation, covered failure, and uncovered failure). Once a component is in the uncovered failure state, the entire system is in that state too as show in Fig.3.

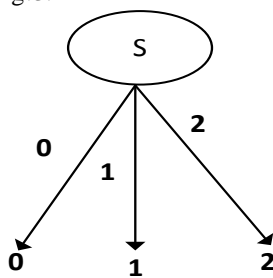

Fig.3 MDD a three-state component of $S$

MDDs have been used in multiple-valued logic design in the circuit area. They have recently been adapted to the reliability analysis of fault tolerant systems in [8].The MDD-based approach is very efficient in determining the severity level of the system, because the MDD is ready and the criteria for severity levels of context indices are also set by models in advance [12] as shown in Fig.4.

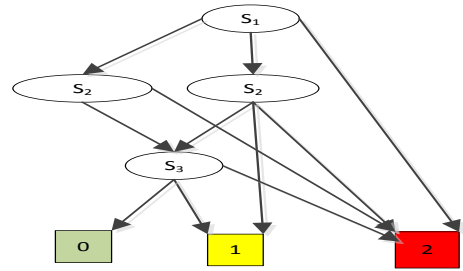

Fig.4 MDD for a severity analysis

Assume that $S$ is a coherent MSS which consists of $n$ components $\{c_1, c_2, \dots, c_i\}$ .The system has $M + 1$ mutual exclusive states $\{0,1,\dots,M\}$, where 0 means that the event component is failed and M symbols the perfect function state of the entire system. Each component $i$ has $M_i + 1$ mutual exclusive states $\{0,1,\dots,M_i\}$, where 0 represents the complete failure state, $M_i$ represents the perfectly functioning state, and $m(0 < m > M_i)$ means that component $C_i$ is in the $(M_i - m)th$ degradation sate. Since these states of components have similar meanings to the variable states in MDD, the modeling method of variables in MDD could be carried out by transforming different events into the states of different variables, as shown in the equation.

$$S_i = \begin{cases} 0, when\ C_{i\ in\ state}\ c_{i0} \\ 1, when\ C_{i\ in\ state}\ c_{i1} \\ \vdots \\ M_i\ when\ C_{i\ in\ state}\ c_{iM_i} \end{cases} \qquad (8)$$

## 3.3 Virtual Sensors Cloud

Sensor-cloud infrastructure provides service instances (virtual sensors) automatically to end users and when requested, in such a way these virtual sensors are part of their IT resources (e.g. disk storage, CPU, memory, etc.) [6]. The virtual sensor enables the use of sensors without worrying about the locations and the specifications of physical sensors. Users can create virtual sensors and freely use them as if they owned the sensors [13]. For example, they can activate or inactivate their virtual sensors, check their status, and set the frequency of resources data collection from them. If multiple users/applications freely control the physical sensors, some inconsistent commands may be issued [10].

## 4. A System Architecture

In this section, we briefly describe the architecture of our proposed framework and present all the steps of diagnosis approach and availability analysis modeling for accelerated fault diagnosis innovation.

## 4.1 A New Hybrid Diagnosis Approach

AFDI provides a new approach and framework for real-time dynamic fault management in cloud computing, addressing the full life cycle of problem determination based on the severity levels and anomalies for VMs and physical server metrics. The AFDI platform is aiming at

IJCSI
www.IJCSI.org

reducing the time and the cost of a fault diagnosis through accelerated fault diagnosis. The AFDI is a new hybrid model using the advantages from multi-valued decision diagram, Naive Bayes classifier intelligent models and a virtual sensors cloud. In this approach we used training data (historical data), as shown in Table.1 the feature metrics collected in our system and applications are from different anomaly symptoms, which consist of records of a set of attributes.

## 4.2 The Architecture of AFDI

The architecture of the proposed AFDI is comprised of four key components, as shown in Fig.5: (a) a monitoring engine which collects and processes the measurement data (reported in Table 2) including basic resource metrics for each virtual machine and physical server; (b) the monitoring database is a data preprocessed module that generates dataset (vectors) after removing outliers and noise; (c) the diagnosis engine with hybrid intelligent models. The first model is the MDD which is used to quickly obtain the overall severity levels of the fault detection and generate an alarm when the fault is serious to a virtual sensor cloud. Otherwise it processes and analyzes an event to execute diagnosis by pre-processing and using training data set of I/O pairs to calculate the parameters estimated by the Naive Bayesian classifier; (d) the virtual sensor cloud receives data from physical sensors and/or alarms from the diagnosis engine, enables provisional service instances to automatically monitor results and signal errors to the system administrator which then can be easily controlled.

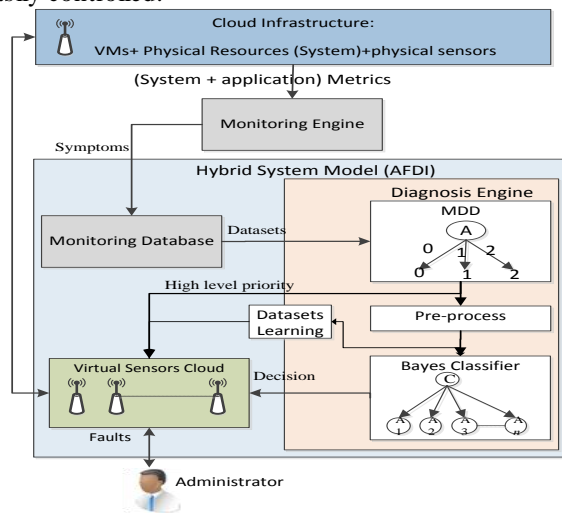

Fig.5 Architecture proposed for a new hybrid system model (AFDI)

Table2: System and application metrics monitored in AFDI

| System Metrics | Description | Measurement Level |
|---|---|---|
| CPU utilization | % CPU time in user-space/ kernel-space | VM, Host |
| Memory usage | % of used memory | VM, Host |
| Network bandwidth | % of Network bandwidth | VM, Host |
| Disk throughput | % of Disk throughput | Host |

| Application Metrics | Description | Measurement Level |
|---|---|---|
| Latency | Response time | VM |
| Throughput | Number of transaction / second | VM |

Here we use a simple example to illustrate the construction process of the multi-state system model of the NBC and MDD. The sub-system consists of a three-state component by NBC model. The system can't work if one of the components fails. We use 0, 1 and 2 to represent the different state of system and components. Where, 0 is a good status (normal working conditions), 1 is a state with minor faults and 2 is the state with serious fault. CPU, memory and network represent three basic events. Node $S$ represents host server state system, as shown in Fig.6.

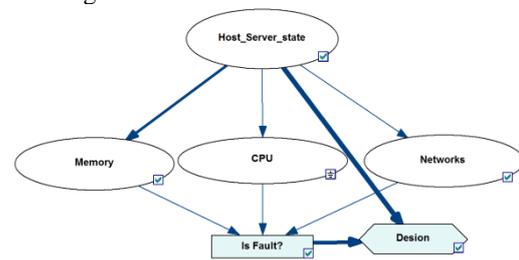

Fig.6 Host server state's model of three-state components

We use bucket elimination algorithm to calculate the probability:

$$P(S) = \sum_{CPU,Mem,Net} P(CPU,Mem,Net,S)$$

$$= \sum_{CPU,Mem,Net} P(S|CPU,Mem,Net)P(CPU)P(Mem)P(Net)$$

$$= \sum_{CPU} P(CPU) \sum_{Mem,Net} P(S|CPU,Mem,Net) \qquad (9)$$

According to the prior probabilities and the conditional probability table (CPT), shown in Fig.7, we can calculate the probability of the system, $P(S)$.

| Parent Node(s) | | | S | | | |
|---|---|---|---|---|---|---|
| Memory | CPU | Network | normal work | minor fault | serious fault | bar charts |
| norml work | normal work | normal work | 0.899 | 0.0685 | 0.0325 | |
| | | minor fault | 0.86 | 0.1039 | 0.0361 | |
| | | serious fault | 0.814 | 0.145 | 0.041 | |
| | minor fault | normal work | 0.698 | 0.211 | 0.091 | |
| | | minor fault | 0.732 | 0.206 | 0.062 | |
| | | serious fault | 0.691 | 0.216 | 0.093 | |
| | serious fault | normal work | 0.641 | 0.208 | 0.151 | |
| | | minor fault | 0.608 | 0.237 | 0.155 | |
| | | serious fault | 0.567 | 0.258 | 0.175 | |
| minor fault | normal work | normal work | 0.515 | 0.279 | 0.206 | |
| | | minor fault | 0.464 | 0.32 | 0.216 | |
| | | serious fault | 0.421 | 0.235 | 0.344 | |
| | minor fault | normal work | 0.361 | 0.371 | 0.268 | |
| | | minor fault | 0.299 | 0.412 | 0.289 | |
| | | serious fault | 0.237 | 0.433 | 0.33 | |
| | serious fault | normal work | 0.175 | 0.454 | 0.371 | |
| | | minor fault | 0.145 | 0.45 | 0.405 | |
| | | serious fault | 0.123 | 0.445 | 0.432 | |
| serious fault | normal work | normal work | 0.125 | 0.309 | 0.566 | |
| | | minor fault | 0.125 | 0.3295 | 0.5465 | |
| | | serious fault | 0.113 | 0.29547 | 0.59153 | |
| | minor fault | normal work | 0.113 | 0.297 | 0.59 | |
| | | minor fault | 0.093 | 0.28331 | 0.62369 | |
| | | serious fault | 0.082 | 0.2728 | 0.6452 | |
| | serious fault | normal work | 0.062 | 0.2717 | 0.6663 | |
| | | minor fault | 0.041 | 0.237 | 0.722 | |
| | | serious fault | 0.0321 | 0.1728 | 0.7951 | |

Fig.7 CPT for subsystem components

In the case of the known information (evidence e); we can generate any query (Q) to get the status of any of the nodes. For example, P(S=Serious-fault/ CPU=serious-fault) or P(S=minor-fault/ Memory=minor-fault) using equation (9) from above we can calculated their probabilities. From the NBC model, we are able to figure out how the components in different states affect the system in different states. Through this analysis we can predict and diagnose the system state to conduct a comprehensive assessment of the availability of the system. This is what traditional availability theory cannot provide.

## 5. Case Studies of Analytic Models

In this section, we discuss quantitative assurance from the analytic models.

### 5.1 IBM BladeCenter Server Host

Blade server systems such as IBM BladeCenter are available and can be used to meet high-availability requirements for many commercial systems [17, 18]. The top-level fault tree for the BladeCenter availability model is shown in Fig.8.

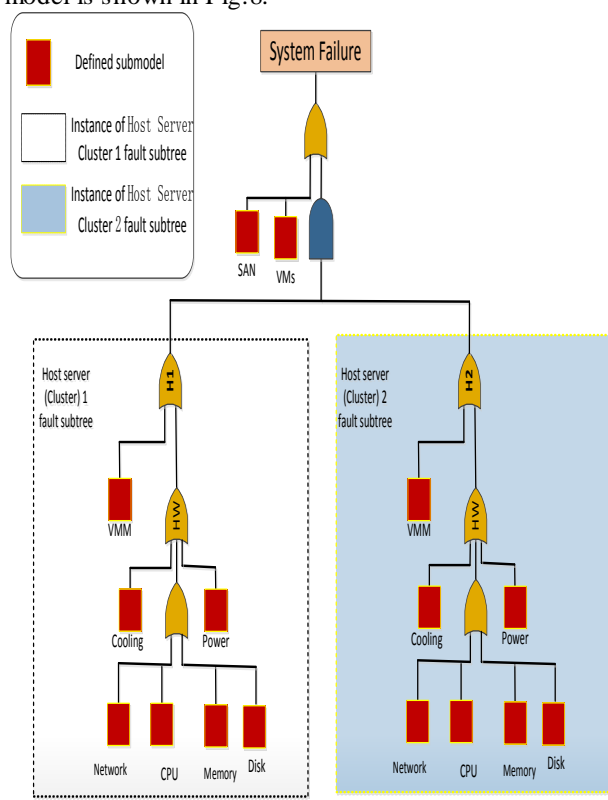

Fig.8 Fault tree for BladeCenter top-level availability model for virtualized cloud platform [18]

The fault tree is a traditional technique to detect and diagnosis faults that are widely employed to generate test sequences for guiding the diagnostic technician. Fig.8 above shows an example of a troubleshooting fault tree for diagnosing a fault in a Blade server system. Actually, this method possesses several intrinsic drawbacks. (1) it is a kind of "if and then" reasoning, and troubleshooting decisions are made based on simple "yes" or "no" judgments. (2) the method has a rigid structure and does not allow the users freedom of choice in terms of their own diagnostic experiences and knowledge; (3) each troubleshooting diagram only deals with a fault case with a single symptom. For a fault case with multiple symptoms, which is common place, the method diagnoses the failures separately; this working mechanism does not allow all failures to be considered in an integrated manner, resulting in a number of unnecessary or ineffective tests and checks [27].

### 5.2 A NBC Availability Models Approach

In this section we convert the FT availability model in Fig.8, into a new approach as the A NBC availability model represented; in Fig.9. A Naive Bayes Classifier is a probability based modeling technique, and it is suitable for knowledge-based diagnostic systems. A NBC enables us to model and reason about uncertainty, ideally suited for diagnosing real world problems where uncertain incomplete data exists [28]. Therefore, it is a suitable solution for fault diagnosis of complex virtualization systems.

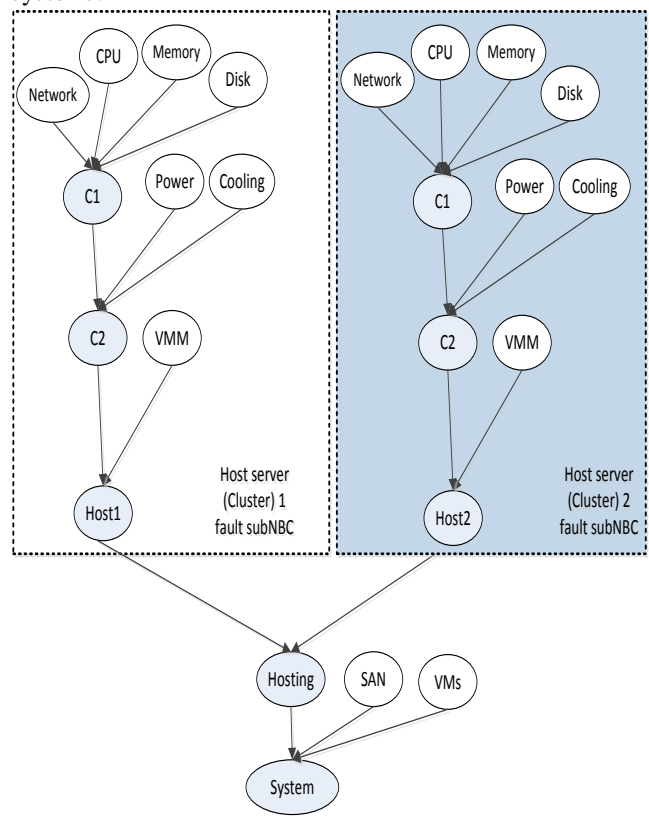

Fig.9 NBC availability model for virtualized cloud platform

A NBC modeling is possible to calculate the posterior probability of the nodes of the network. This gives a measure of the criticality of the component with respect to the occurrence of the fault node (or the top event) and it has been found to have a trend which is very similar to the criticality measure. For example, in a fault diagnostics procedure, when the system is inspected, evidence could be introduced into the components that are found in the working state [22], as shown in Fig.10.

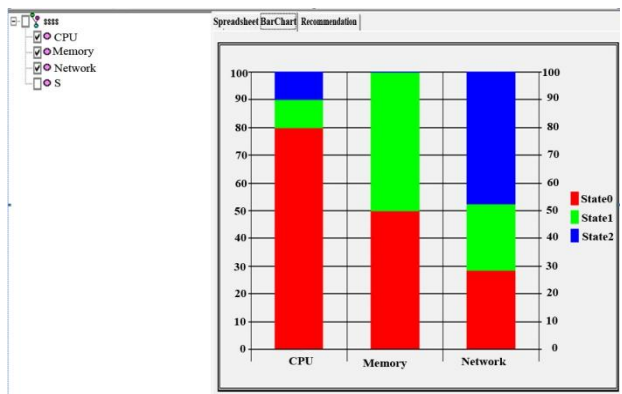

Fig.10 Posterior probabilities with evidences for all components state (MSBNX)

## 5.3 Empirical Evaluation

We applied a first MDD analysis model in our approach model to quickly determine the severity level of the system given a set of detected symptoms as depicted in Fig. 11.

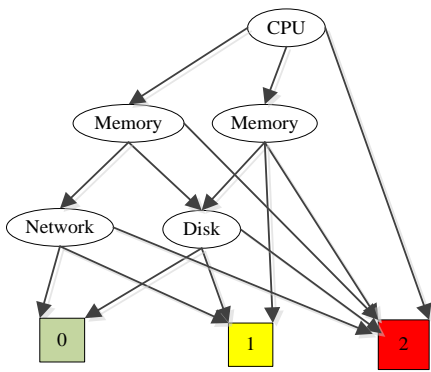

Fig.11 MDD for subsystem severity model analysis approach

For example, if all components (CPU, Memory, Storage I/O and Network) are in good status staying at 0, the severity level of the system is good; if any component stays at status 2 (serious fault), the severity level of the system goes to status 2 and a second model of the Naive Bayes Classifier model begins to generate accelerated fault-diagnosis models to monitor subsystem for three often occurring malfunctions in computing systems (e.g. CPU, memory, and network usages). Their status are key metrics of system, and their abnormal behaviors reflect a state of the system of the inputted into the Naive Bayes Classifier model, where each attribute values belong to a sequence of intervals, as seen in Fig. 6.

For now we will show our simple model on a state based probability model for predicting CPU usage. Its percentage measurement level varies from 0 % to 100 %. We partitioned the measurement level in the intervals. $S_{CPU}$ is a state of CPU usage and $S_{CPU} = \{s_1, ...., s_n\}$ are the different levels of CPU usage. $S_{CPU} \in \{0-25\%, 25-50\%, 50-75\%, 75-100\%\}$ , we observed the percentage of CPU usage at a discrete time $t_1, ...., t_n$, after the classification; we obtain the diagnosis result for probability state of serious fault. $P = \{p_1, p_2, p_3\}$ is a set in which $p_1$ is the probability of "high CPU usage," $p_2$ is the probability of "Memory not enough space," and $p_3$ is probability of "Network overhead", as shown in Fig.12.

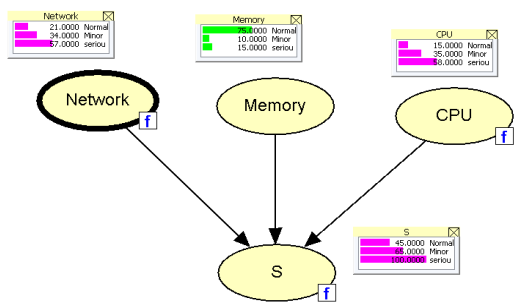

Fig.12 Probability propagation of the faults of NBC model (Hugin software)

## 5.4 Case Study

A virtualization presents an automated and dynamic model, and the physical resources can be dynamically changed when a VM's CPU utilization and network overhead is high for a period of time. The VMM will allocate more resources to the VM and this will result in the CPU utilization and network overhead to decline. Therefore, low throughput in the network with high utilization in CPU can generate an endless loop fault (an infinite loop). Traditional methods cannot detect this type of fault. So, in this case by AFDI, we combine the measurements probability from the VMM level and the physical server resources, so we can form an overall understanding of the system and recognize the fault status rapidly and dynamically.

The structure of the network has been shown in Fig.12. To specify the joint distribution we need to know the prior probability for the variables of the CPU and network components, *(CPU=normal)=0.001, P(CPU=minor)=0.425, P(CPU=serious)=0.573, P(Network=normal)=0.1, P(Network=minor)=0.321 and P(Network=serious)=0.579.*

The CPT for node $S$ (system state) is shown in Fig.7. The posterior probabilities for the components of $S$ (system status) were calculated from the law of probability in Eq. (1) and (3). The results, simulation and performance requirements are shown in Fig.13 and Fig.14.

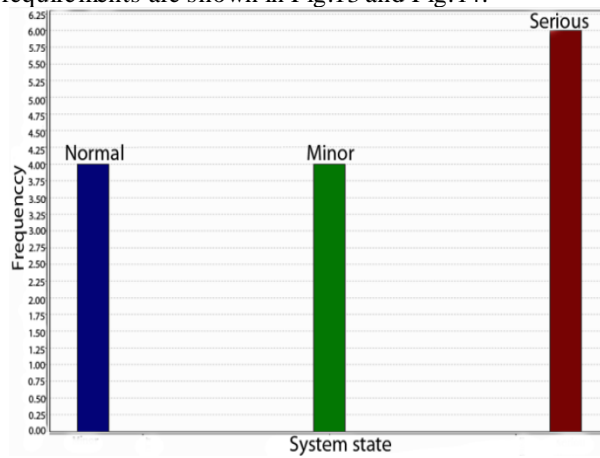

Fig.13 Posterior probabilities of the components, given that the system state has a serious fault (Rapidminor software)

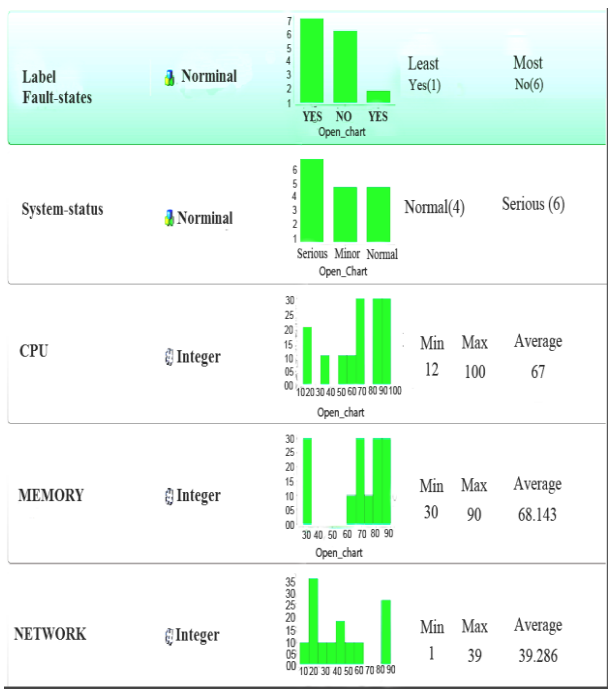

Fig.14 The System and components state simulation results
(Rapidminor software)

## 6. Results and Discussions

Considering in case studies and scenarios, our method's experiment results reveal several interesting findings about MDD and that a Naive Bayes Classifier intelligent models can achieve high accuracy of up to 92.86% detection rate and 6-30% false alarm rate. Performance summary results are shown in Table 3.

Future work will be on how to optimize the using Hadoop MapReduce to speed-up parameters learning in NBC, when wide ranges of input data set size increase.

Table 3: Performance summary results

| Model | Accuracy | Recall | precision | Dataset size(K) |
|---|---|---|---|---|
| AFDI | 92.86% | 90.00% | 95.00% | 800 |
|  | 81.22% | 80.65% | 89.12% | 1900 |

## 7. Conclusion and Future Works

In this paper, we proposed a new model framework for fault diagnosis for cloud computing which is different from the existing methods. To achieve the accelerated fault diagnosis and remediation requirements of accuracy, efficiency and learning ability, a hybrid model (AFDI) tool that combines the MDD and NBC intelligent models with virtual sensors cloud was proposed. AFDI monitors a wide range of metrics across the VMs and the physical server hosting, compared to only non-virtualization resources (e.g. CPU, memory, network and storages), metrics are monitored by most prior work based on the severity levels of consequences according to the symptoms. Therefore, our research increases system availability and the performance of service provider resources. We believe that our work is just a beginning of employing machine learning technologies in IaaS.

Future work includes developing all the functionality for the system. However, much more research and experimentation should be conducted to pioneer future work in this area.

## Appendix

### 1. Hugin Software

Hugin software aims at academics and it allows advanced applications of BNs. The software also incorporates automated learning from databases.
HUGIN software is based on complex statistical models known as Bayesian Networks and influence diagrams. Our modeling technology is used to turn data and expertise into intelligent knowledge management solutions that can reduce fraud losses, improve credit management, lower operational costs, enable legislative compliance and provide more accurate risk assessment [21].

### 2. MSBNx

MSBNx is a free application for the creation and evaluation of BNs. It was used in the initial phases of the research. The figures and the simple calculations of the example in this work are performed with this software. Among the free software's available, MSBNs has been found to be straightforward in the graphical interface and in the probabilities features. However, it has many limitations in particular in terms of the size of the networks that can be created. For this reason, Hugin was chosen for the development of the BNs in this work [26].

### 3. RapidMiner Software

RapidMiner is a software platform developed by the company of the same name. It provides an integrated environment for machine learning, data mining, text mining, predictive analytics and business analytics. RapidMiner is used for business and industrial applications as well as for research, education, training, rapid prototyping, and application development and supports all steps of the data mining process including results visualization, validation and optimization[23].

### Acknowledgments

Thanks Chinese High-tech R&D(863) Program project "Cloud Computer Test and Evaluation System Development (2013AA01A215)" for supporting of this research.

## References

[1] Xu, Fei, et al. "Managing performance overhead of virtual machines in cloud computing: A survey, state of the art, and future directions." Proceedings of the IEEE 102.1 (2014): 11-31.

[2] Barham, Paul, et al. "Xen and the art of virtualization." ACM SIGOPS Operating Systems Review 37.5 (2003): 164-177.

[3] Rish, Irina. "An empirical study of the naive Bayes classifier." IJCAI 2001 workshop on empirical methods in artificial intelligence. Vol. 3. No. 22. IBM New York, 2001.

[4] Nielsen, Thomas Dyhre, and Finn Verner Jensen. Bayesian networks and decision graphs. Springer Science & Business Media, 2009.

[5] Darwiche, Adnan. Modeling and reasoning with Bayesian networks. Cambridge University Press, 2009.

[6] Ponmagal, R. S., and J. Raja. "An Extensible cloud Architecture model for heterogeneous sensor services." International Journal of Computer Science and Information Security 9.1 (2011): 147-155.

[7] Zhai, Sheng, and Shu Zhong Lin. "Bayesian networks application in multi-state system reliability analysis." Applied Mechanics and Materials 347 (2013): 2590-2595.

[8] Xing, Liudong, and Yuanshun Dai. "A new decision-diagram-based method for efficient analysis on multistate systems." Dependable and Secure Computing, IEEE Transactions on 6.3 (2009): 161-174.

[9] Xing, L., Dai, Y.S.: Reliability Evaluation using Various Decision Diagrams. In: 10th Annual International Conference on Industrial Engineering Theory, Applications and Practice, pp. 448–454 (2005)

[10] R. S. Ponmagal and J. Raja, "An extensible cloud architecture model for heterogeneous sensor services," International Journal of Computer Science and Information Security, vol. 9, no. 1, 2011.

[11] Li, Shumin, et al. "A novel decision diagrams extension method." Reliability Engineering & System Safety 126 (2014): 107-115.

[12] Dai, Yuanshun, Yanping Xiang, and Gewei Zhang. "Self-healing and hybrid diagnosis in cloud computing." Cloud computing. Springer Berlin Heidelberg, 2009. 45-56.

[13] Rutakemwa, Maki Matandiko. "From Physical to Virtual Wireless Sensor Networks using Cloud Computing." International Journal of Research in Computer Science 3.1 (2013): 19-25.

[14] Loveland, Scott, et al. "Leveraging virtualization to optimize high-availability system configurations." IBM Systems Journal 47.4 (2008): 591-604.

[15] http://pubs.vmware.com/vsphere-50/index.jsp?topic=%2Fcom.vmware.powercli.cmdletref.doc_50%2FOverview.html

[16] Zaitseva, Elena, and Vitaly Levashenko. "Multiple-Valued Logic mathematical approaches for multi-state system reliability analysis." Journal of Applied Logic11.3 (2013): 350-362.

[17] Smith, W. Earl, et al. "Availability analysis of blade server systems." IBM Systems Journal 47.4 (2008): 621-640.

[18] Trivedi, K., et al. "Availability modeling of SIP protocol on IBM© Websphere©."Dependable Computing, 2008. PRDC'08. 14th IEEE Pacific Rim International Symposium on. IEEE, 2008.

[19] Huang, Yingping, et al. "Bayesian belief network based fault diagnosis in automotive electronic systems." Proceeding of 8th International Symposium on Advanced Vehicle Control. 2006.

[20] Sharma, Bikash, et al. "Cloudpd: Problem determination and diagnosis in shared dynamic clouds." Dependable Systems and Networks (DSN), 2013 43rd Annual IEEE/IFIP International Conference on. IEEE, 2013.

[21] http://www.hugin.com/company-information/about

[22] Lampis, Mariapia. Application of Bayesian Belief Networks to system fault diagnostics. Diss. © Mariapia Lampis, 2010.

[23] http://en.wikipedia.org/wiki/RapidMiner

[24] Zhang, Jianhua, et al. "VMFDF: A Virtualization-based Multi-Level Fault Detection Framework for High Availability Computing." e-Business Engineering (ICEBE), 2012 IEEE Ninth International Conference on. IEEE, 2012.

[25] http://www.saedsayad.com/naive_bayesian.htm

[26] http://research.microsoft.com/en-us/um/redmond/groups/adapt/msbnx/

[27] Huang, Yingping, et al. "Probability based vehicle fault diagnosis: Bayesian network method." Journal of Intelligent Manufacturing 19.3 (2008): 301-311.

[28] Huang, Yingping, Yusha Wang, and Renjie Zhang. "Fault troubleshooting using bayesian network and multicriteria decision analysis." Advances in Mechanical Engineering 2014 (2014).

[29] Zaitseva, Elena, and Vitaly Levashenko. "Multiple-Valued Logic mathematical approaches for multi-state system reliability analysis." Journal of Applied Logic11.3 (2013): 350-362.

[30] Huang, Yingping, Yusha Wang, and Renjie Zhang. "Fault troubleshooting using bayesian network and multicriteria decision analysis." Advances in Mechanical Engineering 2014 (2014).

[31] Smith, W. Earl, et al. "Availability analysis of blade server systems." IBM Systems Journal 47.4 (2008): 621-640.

[32] Mallick, Sayanta, Gaétan Hains, and Cheikh Sadibou Deme. "An Alert Prediction Model for Cloud Infrastructure Monitoring." (2013).

[33] Chandramani Tiwary." Learning Apache Mahout acquire practical skills in Big Data Analytics and explore data science with Apache Mahout", Copyright © 2015 Packt Publishing.

**Ameen Alkasem** is currently a Ph.D. candidate at the school of computer science and technology, Harbin institute of technology, China. He received his M.S. degree from Dalian maritime university in 2010. His research interests include software engineering, fault tolerance computing and cloud computing.

**Dr. Liu Hongwei** focuses on the research and teaching in the field of Computer System Architecture. He has hosted and taken part in over 10 research tasks supported by the government of China. Besides, Dr. Liu also hosted two projects supported by "863"Projects and issue over 90 papers on different journals. He is working on a widely range of subjects, including Parallel Computing and Architecture, Fault Tolerant Computer, Resource allocation and optimization in Cloud Computing System, Evaluation theory and technology in Cloud Computing System, Mobile Computing and Software Reliability Modeling.

**Dr. Zuo Decheng** focuses on the research and teaching in the field of Computer System Architecture, Fault tolerant Computing and Mobile Computing. He has hosted and taken part in over 10 research tasks in the field of fault tolerant computing and mobile computing. Besides, Dr. Zuo also hosted a project supported by "863" Program and issue over 70 papers on different journals. He is working on a widely range of subjects, including Parallel Computing and Architecture, Fault Tolerant Computer, Computer System Architecture Evaluation theory and technology.

**Yao Zhao** received the master's degree in computer science and engineering from HIT, China. He is currently working tow ard the PhD degree at Harbin Institute of Technology(HIT).His research interests include cloud computing, fault-tolerant computing.